\documentclass[nofootinbib,letterpaper,aps,amsmath,amsfonts,twocolumn]{revtex4}
\usepackage[mathscr]{eucal}

\def\eqref#1{(\ref{#1})}

\newcommand{\beeq}{\begin{equation}}
\newcommand{\eneq}{\end{equation}}
\newcommand{\beeqa}{\begin{eqnarray}}
\newcommand{\eneqa}{\end{eqnarray}}
\newcommand{\la}{\langle}
\newcommand{\ra}{\rangle}

\begin{document}
\title{Purifications {\em without} entanglement, State-reconstructions and Entropy}
\author{Chirag Dhara } 
\email{chirag_d4@rediffmail.com}
\affiliation{St. Xavier's College, Mumbai-400001, INDIA}
\author{N.D. Hari Dass} 
\email{dass@imsc.res.in}
\affiliation{The Institute of Mathematical Sciences, Chennai 600-113, INDIA}
\begin{abstract}
In this paper we have analysed in detail
two different purification protocols. The first one, 
proposed by Sudarshan, is based on the {\it preservation of probabilities}.
We have constructed a second protocol here based on {\it optimization of 
fidelities}. 
We have considered both {\it complete} and 
{\it partial} measurements and have established bounds and inequalities 
for various fidelities. For every type of measurement, we have analysed 
post-measurement states based on the Maximum Entropy principle as well as
what we have proposed as unbiased states.
We show that our purification protocol always leads to 
better state reconstruction. These schemes can be thought of as operations in the sense of Kraus and we have explicitly constructed the Kraus operators for these. We have also shown that the entropy either
increases or remains the same depending on the choice of the {\em purification
basis}.
\end{abstract}

\maketitle


\section{Introduction}
{\label{intro}}
In the density matrix formulation \cite{asher} the process of going from a mixed state to a pure state is called 
{\it Purification}.
There is a vast literature on this topic \cite{purerefs}. In the current literature 'Purification' is mostly understood to
be the process of associating a suitable {\it pure} state of a {\em larger}
system whose reduced density matrix is the mixed state one started with.
This necessarily involves {\em entanglement}.
We interpret purification in a larger sense to mean any protocol that produces
a pure state from a mixed state. In this paper we consider purification schemes
{\em without} entanglement.

We specifically consider two schemes for purification the first of which was
introduced by Sudarshan \cite{sudar} based on preservation of probabilities(scheme A), 
and a second proposed by us here based on optimal fidelities(scheme B). We 
apply both these schemes
to the problem of reconstructing the pure state before a measurement from the
post-measurement mixed state. We do so for both partial as well as complete
measurements. In all such cases there is still a lot of freedom in constructing
the post-measurement state itself. We have focussed on two specific choices: The so-called
maximum entropy state $\rho_{max}$ \cite{buz} 
as well as an equal mixture $\rho_{unb}$ of the subensembles 
resulting from
each measurement. We call the latter the 'unbiased state' \cite{ourpaper}. Operationally it
is straightforward to realise as one has to simply divide the initial ensemble
of pre-measurement state into equal parts for each measurement. These are discussed in Sec. IV.A.

We apply our considerations to qubits only. We separately consider the cases
of a) complete measurement involving measurements of $S_x,S_y,S_z$, b)
partial measurement involving measurements of $S_y,S_z$ and finally c) which
involves measuring $S_z$ only. For each of these cases we apply the two
purification schemes mentioned above to both $\rho_{max},\rho_{unb}$. We show
that in all cases scheme-B produces pure states with better overlap with
the unknown initial pure state.

We then show that there exist well defined operations in the sense of Kraus \cite{kra1,kra3}
which realise all these purification schemes. We explicitly construct the
relevant Kraus operators. Though an 'environment' system is needed for
realising these operations, our schemes do not require any entanglement
between the qubit and the environment.

We end our paper with an explicit proof that the total entropy of the qubit
and the environment either remains the same or increases, in accordance with
the second law. Though this is expected, we thought it desirable to explicitly 
prove it.
\section{Some Purification Schemes}
\subsection{Purification Protocol - A} 
 Consider some density matrix which can be regarded as the mixture of two  
orthogonal states 
\beeq
\label{eqn1}
\rho = p_1 \rho_1 + p_2 \rho_2 
\eneq
where, $\rho_{1}^2 = \rho_1, \rho_{2}^2 = \rho_2, tr(\rho_1 \rho_2)=0, tr\rho_1=tr\rho_2 = 1$.

The Purification Protocol discussed here is based on the principle of 
{\it preservation of probabilities}. In \cite{sudar} it was taken to mean
that the {\it overlap} of the {\em purified state} $\rho^{(A)}$ with $\rho_{1,2}$ is $p_{1,2}$. 
 
This Protocol leads to the {\em family} of pure states:
\beeq
\rho^{(A)} =  p_1 \rho_1 + p_2 \rho_2 + \sqrt{p_1 p_2} \frac{\rho_1 \Pi \rho_2 +\rho_2 \Pi \rho_1 }{\sqrt{tr(\rho_1 \Pi) tr(\rho_2 \Pi)}}
\eneq
where $\Pi$ is a projection which is not orthogonal to either $\rho_1$ or $\rho_2$.
If $\rho_1 = |\tilde{0}\ra\la \tilde{0}|,\rho_2 = |\tilde{1}\ra\la \tilde{1}|$ 
and $\Pi$ is of the form $|\chi\ra\la\chi|$ with
$\chi = \mu |\tilde{0}\ra + \nu |\tilde{1}\ra$
($\mu$, $\nu$ $\neq 0$ since $|\chi\ra$ is not orthogonal to either $|{\tilde0}\ra
$ or 
$|{\tilde 1}\ra$) then the purified state is given by
($\phi$ is the phase of $\mu \nu^{\star}$) 
\beeq
\label{purified}
\rho^{(A)} = |\psi\ra\la\psi|;~~~~|\psi\ra = \sqrt{p_1}~|\tilde{0}\ra+\sqrt{p_2}e^{-i\phi}~|\tilde{1}\ra 
\eneq

The reason that only this 
phase appears in the purified $\rho$ is that preservation of probabilities 
leaves only a phase left unspecified in a pure state. Different choices of 
$\phi$ lead to different purified states. There is no principle that selects 
a particular value of $\phi$. This protocol can only be implemented
probabilistically.

\subsection{Purification Protocol - B}
Consider the state
\beeq
\label{raw}
\rho =  \left( \begin{array}{cc}
a & p \\
p^\star & 1-a
\end{array} \right) 
\eneq
which we wish to 'purify'.
The principle we adopt to fix 
the purified state is: {\it the purified 
state must have maximal overlap with the mixed state we started with}. (The overlap between states $\rho_1$ and $\rho_2$ is
${\cal F} = tr (\rho_1. \rho_2)$).
This amounts to demanding that the purified state be {\it as close as possible} 
to the mixed state. In contrast to A this is a deterministic protocol. 

%
It is elementary to show that the pure state $\rho_\psi$ with maximum overlap
with $\rho$ is the eigenstate of $\rho$ corresponding to the maximum
eigenvalue \footnote{ We thank the referee for pointing this out to us.}. To see
this let $\lambda_+$ be the maximum eigenvalue(considered degenerate for
simplicity) and let the others be $\lambda_i, i=1,...,N-1$ where $N$ is the dimensionality of
the Hilbert space, and let $|\lambda\rangle$ be the corresponding eigenstates.
Denoting the pure state by $|\psi\rangle = \alpha_+|\lambda_+\rangle + \sum_i
\alpha_i|\lambda_i\rangle$ we have
\beeq
tr \rho\rho_\psi 
= \lambda_+ +\sum_i |\alpha_i|^2(\lambda_i-\lambda_+)
\eneq
Since $\lambda_i-\lambda_+$ is negative it is obvious that the overlap is
maximised when $\alpha_i = 0$ i.e $|\psi\rangle = |\lambda_+\rangle$ and
that the maximum value of the overlap equals $\lambda_+$.
It is
easy to show that 
\beeq
\lambda_+ = a+\Delta/2;~~\Delta = 1-2a+\sqrt{4|p|^2+(1-2a)^2}
\eneq
and
\beeq
\label{maxover}
\rho_\psi = {2\over {\Delta(\Delta+2a-1)}}  \left( \begin{array}{clcr}
|p|^2 & p\Delta/2 \\
p^*\Delta/2 &\Delta^2/4 
\end{array} \right) 
\eneq

\section{Kraus Formalism and Purification.} 
Kraus {\it et al} \cite{kra1,kra3} have given a formalism to study all possible 
changes of quantum states through the so-called operations. 
An {\it Operation} $O$ is defined as follows: Consider a quantum system in 
the state $\rho_{sys}$ with a Hilbert Space ${\cal H}$ which is coupled to 
another quantum system, often called the {\it environment}, in the state 
$\rho_{E}$ 
and which has a state space ${\cal H}_{E}$. The system and the 
environment 
interact through a Unitary Evolution $U$ which acts on the total Hilbert Space 
${\cal H} \otimes {\cal H}_E$. Some property of the environment is 
{\it selectively} measured (what this means is that one selects a particular
measurement outcome), represented by a projection operator $Q_E$ 
so that the combined state becomes:
\beeq
\label{fund}
\hat \rho=({\bf I} \otimes Q_E) U (\rho_{sys} \otimes \rho_E)U^\dagger ({\bf I} \otimes Q_E)  
\eneq
The system is then described by the 
{\it reduced density matrix},
$\hat \rho_{sys}={Tr}_{E}~~\hat \rho$;
where the trace is taken over all possible states of the environment.
The resulting state change 
$O:\rho_{sys} \longrightarrow \hat \rho_{sys}$ 
is called an Operation.
This operation can be represented 
in terms of \emph{ Kraus operators} ${\cal A}_{k}$ acting on the state space 
of the system such that
\beeq
\label{kreln}
\hat \rho_{sys}=\sum_{k \epsilon K} {\cal A}_{k} \rho_{sys} {\cal A}_{k}^\dagger
\eneq
The operators ${\cal A}_{k}$ are defined by 
\beeq
\label{defn}
(f, {\cal A}_{k} g)=((f \otimes f_{k}^{E}), U (g \otimes g^{E}))
\eneq
where $f$, $g$ arbitrary vectors in the state space of the system, 
$\{f_{k}^{E}~ | k \epsilon K \}$ are an orthonormal basis of 
$Q_{E}{\cal H}_E$ {\em extended} to ${\cal H}_E$ and $g^{E}$ is the pure state in which the environment 
can be assumed to have started in.
As the measurement $Q_E$ is {\it selective}, the ${\cal A}_{k}$ operators satisfy 
the {\it trace non-increasing} condition
$\sum_{k \epsilon K} {\cal A}_{k} {\cal A}_{k}^\dagger \leq {\bf I}$
where $K$ is some indexing set.

Operations can connect any given pair of density 
matrices $\{\rho_{1}, \rho_{2}\}$, and in particular, a mixed state and a pure state.
The entropy aspects of this are discussed in section V. 

\subsection{Kraus Operators for Qubits}
We shall relax the condition 
of selectivity in measurements and consider all possible outcomes for 
measurements.
We 
shall also restrict ourselves to 
qubits only. Then one needs two Kraus operators for a general operation.

Consider {\em some} orthonormal basis (called purification basis) $|0\ra,|1\ra$; and any pair of operators
\beeq
\label{koperators}
{\cal A}_{0}= |\psi\ra\langle0| ~~~ 
{\cal A}_{1}= |\psi\ra\langle1| ~~~
\eneq
satisfying 
$A_{0}^\dagger {\cal A}_{0} + {\cal A}_{1}^\dagger {\cal A}_{1} = {\bf I}$.
For any arbitrary density matrix $\rho_{in}$, these operators produce
\beeq
\rho_{out}={\cal A}_{0} \rho {\cal A}_{0}^\dagger + {\cal A}_{1} \rho {\cal A}_{1}^\dagger 
= |\psi\ra\la\psi|
\eneq
Clearly, $\rho_{out}$ is a pure state and it 
is {\it independent} of the initial state $\rho_{in}$.
By eqn(\ref{defn}) the Kraus operators here are of the form,
\beeq
{\cal A}_{0}= \langle0_E| U |0_E\rangle~~~
{\cal A}_{1}= \langle1_E| U |0_E\rangle
\eneq
where the initial state of the environment is the pure state $|0_E \ra$.
It is straight forward to check that the  unitary operator $U$ that generates 
the Kraus operators of eqn(\ref{koperators}) is (with the notation
$|\psi,\chi\ra = |\psi\ra|\chi\ra_E,\la\xi|\psi\ra = 0$):
\beeq
\label{Unitary}
U 
= \sum_{s=0}^{1}(|\psi,s\ra\la s,0| + |\xi,s\ra\la s,1| )
\eneq
Kraus Operators relevant for the purification schemes of Sec.II can easily be worked out.

\section{ State reconstruction for qubits and efficiency of purifications.
}
Now we apply these considerations to address the 
issue of {\em optimal reconstruction} of {\em unknown} quantum states through
both complete as well as partial measurements. The question we answer is:
'which of these schemes produces a pure state that is closest to the pre-measurement pure state?'
\subsection{Post measurement state}

Clearly there are many ways of performing both complete as well as partial
measurements leading to different post-measurement (mixed) states. Here we
consider two particular ways of obtaining the post-measurement state: i) first
is based on the well known {\em maximum entropy principle} \cite{maxent} and we shall
denote the corresponding post-measurement state by $\rho_{max}$. Such states
have been studied extensively in \cite{buz}; ii) second is a proposal we are
making in this paper which we call the {\em unbiased mixture} \cite{ourpaper} wherein the
mixed states arising from different measurements are equally mixed to give
the post-measurement state $\rho_{unb}$. Since the first is discussed at length in \cite{buz} we briefly describe the second.

Consider an ensemble containing
N identical copies of some unknown quantum state described by the pure density
matrix $\rho_{ini}$. One then subdivides this ensemble into $k$ identical
subensembles, makes measurement of $S_z$(equivalently $S_y$ or $S_x$) if
$k=1$, makes measurements of $S_z$ {\em and} $S_y$ (equivalently $S_y$ and
$S_x$ etc..) if $k=2$, and finally measures all of $S_x,S_y,S_z$ if $k=3$.
One then puts together the resulting mixed ensembles in each case to form
the {\it post-measurement mixed state} $\rho_{unb}$.

Now we construct these post-measurement states and apply the two purification
schemes to each case.

We use the eigenstates of $\sigma_{z}$, $|\pm\ra_{z}$ as the basis for the Hilbert space. The eigenstates of $\sigma_{x}$, $\sigma_{y}$ are given by $|\pm\ra_{x} = \frac{1}{\sqrt{2}}(|+\ra_{z} \pm |-\ra_{z})$; $|\pm\ra_{y} = \frac{1}{\sqrt{2}}(|+\ra_{z} \pm i|-\ra_{z})$.
\subsection {Complete measurement: $S_{x}, S_{y}, S_{z}$ measured.}
Let $p_{1}$ be the probability for the outcome  
$|+\ra_{z}$, $p_{2}$ for 
$|+\ra_{y}$ and $p_{3}$ for $|+\ra_{x}$.
The post-measurement density matrix for the $S_z$ measurement is:
\beeq
\rho_{1} = p_{1}|+\ra_{zz}\la+| + (1 - p_{1}) |-\ra_{zz}\la-|
\eneq
and similarly for the other two subensembles.
Now one takes an equal weightage of the three post-measurement density
matrices to give
$\rho_{unb,3} = (1/3) (\rho_{1}+\rho_{2}+\rho_{3})$. 
With the notation
$A_j=2p_j-1; j=1,2,3$ we can write,
\beeq
\rho_{unb,3} = {1\over 6}\left( \begin{array}{cc}
A_1+3 & A_3-iA_2 \\
A_3+iA_2 & 3-A_1
\end{array} \right) 
\eneq
where '3' denotes the number of orthogonal spin components measured. Knowing 
$p_1,p_2,p_3$ amounts to a {\it complete} measurement and the initial density 
matrix has to be 
\beeq
\label{inistate}
\rho_{ini}= {1\over 2}\left( \begin{array}{cc}
1+A_1 & A_3-iA_2 \\
A_3+iA_2 & 1 - A_1
\end{array} \right) 
\eneq
Clearly, the relation between $\rho_{unb,3}$ and $\rho_{ini}$ is, 
\beeq
\label{msmt}
\rho_{unb,3} = (1/3) ({\bf I} + \rho_{ini})
\eneq
In an earlier publication \cite{ourpaper} we have established the result analogous to 
eqn(\ref{msmt}) for {\em arbitrary} systems with finite dimensional ${\cal H}$ 
. 
The Fidelity of $\rho_{unb,3}$ with the initial state is:
\beeq
\label{fid1}
{\cal F} (\rho_{ini},\rho_{unb,3}) = tr (\rho_{ini} \rho_{unb,3}) 
= 2/3
\eneq
independent of $\rho_{ini}$.
Since, $\rho_{ini}$ is pure, its  eigen-values  are $0$, $1$. Hence, 
the eigenvalues of $\rho_{unb,3}$ are, from (\ref{msmt}), $1/3$ and $2/3$. 
Therefore,
\beeq
\label{111}
\rho_{unb,3} ={2\over 3}|l\ra\la l|+{1\over 3}|s\ra\la s|
\eneq
Substituting in eqn(\ref{msmt}) and using the completeness relation 
${\bf I} = |l\ra \la l|+ |s\ra \la s|$, 
one finds
\beeq
\label{inist}
\rho_{ini}=|l\ra\la l|
\eneq
The eigenvector of $\rho_{unb,3}$ corresponding to the 
{\em largest} eigenvalue is the initial state.
The Purification of (\ref{111}) by Protocol - A is (see eqn(\ref{purified}))
\beeqa
\rho^{(A)}_{unb,3} &=&
 (2/3)  |l\ra \la l| + (1/3)  |s\ra \la s|\nonumber\\ 
 &+&  \sqrt{2}/3 (e^{i \phi}   |l\ra \la s|
 + e^{- i \phi} |s\ra \la\ l|)
\eneqa
This is clearly not the initial state.
In fact, the fidelity of this state with the initial pure state is still only,
\beeq
{\cal F}(\rho_{ini},\rho^{(A)}_{unb,3}) = 2/3
\eneq

On the other hand, the purification of $\rho_{unb,3}$ by protocol- B is simply to select the state with the largest eigenvalue i.e. $$\rho_{unb,3}^{(B)}=|l\ra\la l|$$
which is indeed the initial pure state by virtue of eqn(\ref{inist}).
\emph{The initial state is the closest pure state 
to the mixed state $\rho_{unb,3}$}
Protocol - B has purified the post-measurement state $\rho_{unb,3}$
to maximum fidelity(in this case 100\%) with the initial state. 

State reconstruction by the MaxEnt Principle \cite{buz,maxent} gives:
\beeq
\rho_{max,3} = {1\over 2}({\bf I} + A_1\sigma_{z} + A_2\sigma_{y} + A_3\sigma_{x})
= \rho_{ini}
\eneq
Clearly, in the case of a complete measurement state reconstruction by the MaxEnt Principle and purification by protocol-B both yield the initial state.

It should also be emphasised that even if one did not know that $p_i$ were
the probabilities of outcome from complete measurements, but only knew the form
of $\rho_{unb,3}$, B would nevertheless reconstruct the initial state. 
The power of B is  clearly evident in the case of partial measurements, discussed next .

\subsection{Partial Measurements: $S_{y}, S_{z}$ measured.}
In the case of partial measurements 
the initial state {\it can never} be unambiguously reconstructed.
In this section we establish the following two results: (i) the
purified state under protocol-B {\em always} has a greater fidelity
with the initial state than does the post-measurement state, (ii)
the fidelity of the purified state under protocol-B with the initial
state is {\em always} greater than that of the purified state under
protocol-A (in an phase-average sense as protocol-A does not {\em favour}
any single pure state) except in some singular cases where the 
fidelities are the same. Thus protocol-B is the better when trying
to reconstruct the initial state from the post-measurement
state.

Suppose two measurements are made. Let $p_{1}$ and $p_{2}$ by the probabilities of outcome for $|+\ra_{z}$ and $|+\ra_{y}$ respectively.
Then post-measurement state (in the notation defined in Sec.IV.B) is, 
\beeq
\label{part}
\rho_{unb,2} = \frac{1}{4}\left( \begin{array}{cc}
A_1+2 & -iA_2 \\
iA_2 & 2-A_1
\end{array} \right)
\eneq
Let, the initial state be $\psi_{ini} = \alpha|+\ra_{z} + \beta|-\ra_{z}$.
Then, $(1+A_1)/2 = |\alpha|^2$, $Im(\alpha\beta^*) = A_2/2$.
From these relations we can compute the fidelity:
\beeq
\label{F1}
{\cal F}(\rho_{ini}, \rho_{unb,2}) = \frac{1}{4}(A_{1}^{2} + A_{2}^{2} + 2)
\eneq
where $\rho_{ini} = |\psi_{ini}\ra\la\psi_{ini}|$. 

To purify the state by protocol - A, we can adopt the following procedure: 
From (\ref{part}) we find the eigenvalues of $\rho_{unb,2}$ to be $p_1 = (2 + |A|)/4$ and $p_2 = (2 - |A|)/4$, where $|A| = (A_1^2 + A_2^2)^{1/2}$.
Using the notation ${\bf A} = (A_1, A_2)$ and ${\sigma} = (\sigma_z, \sigma_y)$ we can write $\rho_{unb,2}$ as a mixture of two orthogonal pure states:
\beeq
\label{diagonal}
\rho_{unb,2} = p_1(\frac{{\bf I}}{2} + \frac{\hat {\bf n}.{\bf \sigma}}{2}) + p_2(\frac{{\bf I}}{2} - \frac{\hat {\bf n}.{\bf \sigma}}{2})
\eneq
where ${\bf A} = |A|\hat {\bf n}$. With $\rho_1 = (\frac{{\bf I}}{2} + \frac{\hat {\bf n}.{\bf \sigma}}{2})$ and $\rho_2 = (\frac{{\bf I}}{2} - \frac{\hat {\bf n}.{\bf \sigma}}{2})$, this is of the form (\ref{eqn1}).

The initial state from (\ref{inistate}) is, 
\beeq
\label{initial}
\rho_{ini} = \frac{{\bf I}}{2} + \frac{{\bf A}.{\bf \sigma}}{2} \pm \frac{1}{2} \sqrt{1 - |A|^2} \sigma_x
\eneq
Denoting $\rho_1 = |+'\ra \la +'|$ and $\rho_2 = |-'\ra \la -'|$ the purification of $\rho_{unb,2}$ by protocol-A is,
\beeqa
\label{rhoproA}
\rho_{unb,2}^{(A)} &=& (\frac{1}{2} + \frac{|A|}{4})|+'\ra \la +'| + (\frac{1}{2} - \frac{|A|}{4}) |-'\ra \la -'| \nonumber\\
&+& \sqrt{\frac{1}{4} - \frac{|A|^2}{16}} (|+'\ra \la -'|e^{i\beta} + |-'\ra \la +'|e^{-i\beta})
\eneqa
Then
\beeq
tr(\rho_{unb,2}, \rho_{unb,2}^{(A)}) = \frac{1}{2} + \frac{|A|^2}{8}
\eneq
The eigenstates of $\rho_{unb,2}$ are:
\beeq
|\pm'\ra^T = {N_\pm} (iA_2, A_1 \mp |A|)
\eneq
where $N_\pm^{-2} = 2(|A|^2 \mp |A.A_1|)$ are the normalization factors.
Using $N_+.N_- = 2|A.A_2|$ and $\la +'|\sigma_x|-'\ra =  -ie^{i\delta}$
( $\delta$ is the phase of $A_2$) and  
$\la \pm'|\sigma_x|\pm'\ra =0$
we can compute
${\cal F}(\rho_{ini}, \rho_{unb,2}^{(A)})$. 
This is a function of the arbitrary phase 
$\beta$ (from protocol-A). 
We therefore average the fidelity over the $\beta$. This can be done 
in two ways: i) Average over all possible values of the $\beta$ and ii) Average 
over only those values of the $\beta$ (two such values) which gives 
optimal fidelity with the two possibilities for the initial state
(eqn(\ref{initial})). After some tedious work one gets
\beeq
{\cal F}_{av}(\rho_{ini},\rho^{(A)}_{unb,2}) = 
{\cal F}(\rho_{ini}, \rho_{unb,2})
\eneq
Using eqn(30) and $|A|\le 1$ one sees that this fidelity can never exceed $5/8$ for any initial state.
Now, it can be verified that purification by protocol-B (\ref{maxover}) gives,
\beeq
\rho^{(B)}_{unb,2}= \left( \begin{array}{cc}
\tilde{p} & -ie^{i\delta}\sqrt{\tilde{p} (1- \tilde{p})}  \\
ie^{-i\delta}\sqrt{\tilde{p} (1- \tilde{p})} & 1 - \tilde{p}
\end{array} \right)   
\eneq

where

\beeq
\tilde{p} = {1\over 2}(1+{\hat n}_1 )
\eneq
Using the relations for $|\alpha|^2$, $A_{1}$ and $A_{2}$,
\beeq
\label{fid3}
{\cal F}(\rho_{ini}, \rho^{(B)}_{unb,2}) =  \frac{1}{2} (1 + |A|) 
\eneq
This can indeed reach unity for a large class of initial states. From (\ref{fid3}),(\ref{F1}) it is clearly seen that, 
\beeq
{\cal F}(\rho_{ini}, \rho_{unb,2}^{(B)}) \geq {\cal F}(\rho_{ini}, \rho_{unb,2})
\eneq



State reconstruction by the MaxEnt Principle \cite{maxent} leads to:
\beeq
\label{ME2}
\rho_{max,2} = \frac{1}{2} \left( \begin{array}{cc}
1+A_1 & -iA_2  \\
iA_2 & 1-A_1
\end{array} \right)   
\eneq
and we have the relation:
\beeq
\label{rel}
\rho_{unb,2} = \frac{1}{4}({\bf I} + 2\rho_{max,2})
\eneq
Using this relation we get the fidelity:
\beeq
\label{fidmax2}
{\cal F}(\rho_{ini}, \rho_{max,2}) = \frac{1+|A|^2}{2}
\eneq
Relation (\ref{rel}) implies that the eigenstates of $\rho_{max,2}$ are the same as that for $\rho_{unb,2}$. Further, the eigenvalues can also be calculated directly from (\ref{rel}). They are $q_1 = (1 + |A|)/2$ and $q_2 = (1 - |A|)/2$.
Purification of $\rho_{max,2}$ by protocol-A gives,
\beeqa
\rho_{max,2}^{(A)} &=& (\frac{1 + |A|}{2})|+'\ra \la +'| + (\frac{1 + |A|}{2}) |-'\ra \la -'| \nonumber\\
&+& \sqrt{\frac{1 - |A|^2}{4}} (|+'\ra \la -'|e^{i\beta} + c.c.)
\eneqa
Again, using similar arguments as earlier in this section, we can calculate:
\beeq
{\cal F}_{av}(\rho_{ini}, \rho_{max,2}^{(A)}) = 
{\cal F}(\rho_{ini},\rho_{max,2})
\eneq
Since the largest eigenvalue eigenstate of $\rho_{max,2}$ is the same as that for $\rho_{unb,2}$, we have $\rho_{max,2}^{(B)} = \rho_{unb,2}^{(B)}$. 
Finally we have,
\beeqa
& &{\cal F}(\rho_{ini}, \rho_{max,2}^{(B)}) = {\cal F}(\rho_{ini}, \rho_{unb,2}^{(B)})\nonumber\\ 
&\geq& {\cal F}(\rho_{ini}, \rho_{max,2}) =  {\cal F}_{av}(\rho_{ini}, \rho_{max,2}^{(A)})\nonumber\\ 
&\geq& {\cal F}(\rho_{ini}, \rho_{unb,2}) = {\cal F}_{av}(\rho_{ini}, \rho_{unb,2}^{(A)})
\eneqa
\subsection{Partial Measurement: $S_{z}$ measured.}
When only one component of spin is measured, say, $S_{z}$ we have
$$\rho_{unb,1} = p_{1} |+\ra_{zz} \la+| + (1 - p_{1})|-\ra_{zz} \la-|$$
Again, the initial state is of the form: $\psi_{ini} = \alpha |+\ra_{z} + \beta |-\ra_{z}$. Now, we know only that $|\alpha|^2 = p_{1}$. 
The Fidelity of the post-measurement state with the initial state is:
\beeq
{\cal F}(\rho_{ini}, \rho_{unb,1}) = p_{1}^2 + (1 - p_{1})^2 \nonumber\\
\eneq

Now, if we purify $\rho_{unb,1}$ by protocol- A, then
\beeq
\rho_{unb,1}^{(A)} = \left( \begin{array}{cc}
p_{1} & \sqrt{p_{1}(1 - p_{1})}e^{i \phi} \\
\sqrt{p_{1}(1 - p_{1})}e^{-i \phi} & 1 - p_{1}
\end{array} \right)
\eneq
$\rho_{ini}$ is of the form:
\beeq
\rho_{ini} = \left( \begin{array}{cc}
p_{1} & \sqrt{p_{1}(1 - p_{1})}e^{-i \theta} \\
\sqrt{p_{1}(1 - p_{1})}e^{i \theta} & 1 - p_{1}
\end{array} \right)
\eneq
where $p_{1}$ is known from measurement, but the phase $\theta$ is unknown.
The average Fidelity with equal weightage for all $\theta$ is,  
\beeq
{\cal F}_{av}(\rho_{ini}, \rho_{unb,1}^{(A)}) = p_{1}^2 + (1 - p_{1})^2 
\eneq
Now, if $p_{1}$ $\geq$ $1/2$, then the purified state according to protocol-B is,
\beeq
\rho^{(B)}_{unb} = |+\ra_{zz}\la+|
\eneq
Then,
\beeq
{\cal F}(\rho_{ini}, \rho^{(B)}_{unb,1}) = p_{1}
\eneq
Since, $p_{1}$ $\geq$ $1/2$, it can be verified that 
\beeq
{\cal F}(\rho_{ini}, \rho_{unb,1}^{(B)}) \geq {\cal F}(\rho_{ini}, \rho_{unb,1}) 
= {\cal F}_{av}(\rho_{ini}, \rho_{unb,1}^{(A)})
\eneq
It can be verified that even for $p_{1}$ $<$ $1/2$ protocol- B always leads to an improvement in fidelity.
\emph{Thus, the fidelity offered by protocol-B is better than the average 
fidelity offered by protocol- A.}

In this case the state reconstruction by the MaxEnt Principle is:
\beeq
\rho_{max,1} = \left( \begin{array}{cc}
p_{1} & 0 \\
0 & 1 - p_{1}
\end{array} \right)
= \rho_{unb,1}
\eneq
Hence, the fidelity calculations remain the same as for $\rho_{unb,1}$.

\section{Entropy of purification.}
As mentioned before, the purification process leads to interesting questions
regarding entropy. Associated with every quantum state is the von Neumann
entropy
\beeq
S_v = - tr \rho\ln\rho
\eneq
In terms of the eigenvalues $\{\lambda_i\}$ of the density matrix, one has
\beeq
S_v = -\sum_i~\lambda_i~\ln \lambda_i
\eneq
Hence the entropy $S_v$ is zero for pure states. On the other hand, for
mixed states this entropy is always positive. Thus purification 
reduces the entropy of the  system. But the real issue is what 
purification, understood as an Operation on the composite of the 
system and the environment, does to the total entropy.

To address this, let us first consider the action of the unitary operator
$U$ in eqn(\ref{fund}) which is given explicitly in eqn(\ref{Unitary}). The
initial state of the composite is $\rho^{(0)}\otimes |0_E\ra\la0_E|$. It
is easy to see that
\beeq
\label{ent1}
U(\rho^{(0)}\otimes|0_E\ra\la0_E|)U^\dag = |\psi\ra\la\psi|\otimes \rho_E^{'}
\eneq
where $\rho_E^{'}$ is given by
\beeqa
\label{ent2}
\rho_E^{'}& =& \rho^{(0)}(0,0)|0_E\ra\la0_E|+\rho^{(0)}(1,0)|1_E\ra\la0_E|\nonumber\\
&&+\rho^{(0)}(0,1)|0_E\ra\la1_E|+\rho^{(0)}(1,1)|1_E\ra\la1_E|
\eneqa
There are two remarkable properties of eqns(\ref{ent1},\ref{ent2}): i) the
unitary transformation maintains the disentangleness of the system and
environment and ii) the von Neumann entropy of the environment, which was
zero to start with, has become equal to the system entropy we started
with. So at this stage, not surprisingly, the total entropy has not
changed(unitary transformations cannot change this!). This is an
example of {\it entropy swapping}.

Now we consider the effect of the projections $Q_E$. Remembering that for a
total and not selective measurement we need two of them, the
resulting environment density matrix is given by
\beeq
\label{entfin}
\rho_E^{fin} = \rho^{(0)}(0,0)|0_E\ra\la0_E|+\rho^{(0)}(1,1)|1_E\ra\la1_E|
\eneq
If the {\em purifying basis} $|0\ra,|1\ra$ is the same as the eigenbasis
of $\rho^{(0)}$, it is immediately obvious that the entropy of $\rho_E^{fin}$
is the same as that of $\rho^{(0)}$ and the purification process does
not change the total entropy.

If that is not so, the entropy of $\rho_E^{fin}$ is still the entropy of
a density matrix $\rho^{(2)}$ which is diagonal and whose 
{\em diagonal} elements are the same as that of $\rho^{(0)}$. It
therefore follows that the determinant of $\rho^{(2)}$ is greater than
that of $\rho^{(0)}$. Now we establish an important result: {\it the von
Neumann entropy of a two level system is an increasing function of
the determinant of
the density matrix.}

To prove this, note that the eigenvalues of the density matrix are given
by $\lambda_\pm = (1\pm\sqrt{1-4\Delta_\rho})/2$ where $\Delta_\rho$ is the determinant
($ 0\le\Delta_\rho\le 1/4$). 
It is then straight forward to show that
\beeq
{\partial S_v\over\partial\Delta_\rho} = {1\over\sqrt{1-4\Delta_\rho}}\ln{1+\sqrt{1-4\Delta_\rho}\over 1-\sqrt{1-4\Delta_\rho}}~\ge 0
\eneq
In this case the total entropy {\em increases}.

\section{Conclusion}
In this paper we have addressed schemes for producing pure states from mixed
states. Though our schemes necessarily involve an environment, what is novel
is that no entanglement between the system and environment is necessary. In this
respect our purification schemes differ from those that are mostly dealt with in the current literature. We give explicit expressions for the Kraus operators
that are required.

We have considered two purification schemes A and B
as well as two post-measurement states, one of which is based on the maximum 
entropy principle. In all cases considered we find that scheme B purifies
more efficiently than scheme A. 
The unbiased mixture state 
and the MaxEnt states perform equally well in all cases in 
producing a
pure state as close as possible to the pre-measurement state. 
Operationally, unbiased mixture states are straight forward to realise 
compared to MaxEnt states.

Finally we show that the total entropy either increases or remains the same.
This is possible because we are able to achieve purifications
without entanglement resulting in a case of entropy swapping. 
This along with a theorem we prove on entropy of two-state systems leads to the result.

\section{Acknowledgements}
CD thanks Prof. Ajay Patwardhan of St. Xavier's College, Mumbai, 
for his continuous encouragement and guidance. CD thanks The Institute of 
Mathematical Sciences for a Summer Student Fellowship during 2004 and the
Indian Academy of Sciences for a fellowship in 2006.

\end{document}